\documentclass{elsart}
\usepackage{graphicx}% Include figure files
\usepackage{epsfig}
\usepackage{dcolumn}% Align table columns on decimal point
\usepackage{bm}% bold math
\newcommand{\ud}{\mathrm{d}}

\begin{document}
\begin{frontmatter}

\title{Magnetic horizons of UHECR sources and the GZK feature} \author{Olivier
Deligny$^a$, Antoine Letessier-Selvon$^a$, Etienne Parizot$^b$}
\address{$^a$ LPNHE, Universit\'es de Paris 6 et 7 , C.N.R.S., 4
place Jussieu T33 RdC, F-75252 Paris Cedex 05, France\\
$^b$ IPN Orsay, Universit\'e de Paris 11, C.N.R.S., Orsay Cedex, France}

\begin{abstract}
We study the effect of random extra-galactic magnetic fields on the
propagation of protons of energy larger than $10^{19}$ eV. We show
that for reasonable field values (in the 100~nG range) the transition
between diffusive and ballistic regimes occurs in the same energy
range as the GZK cutoff (a few $10^{19}$ eV).  The usual
interpretation of the flux reduction above the GZK energy in terms of
a sudden reduction of the visible horizon is modified.  Moreover,
since the size of the diffusion sphere of a continuous source of
cosmic rays is of the order of 10 Mpc, the local structure of the
Universe and therefore of potential local astrophysical sources plays
a dominant role in the expected spectrum.  Under reasonable
assumptions on the sources configurations the expected GZK cutoff is
reduced.
\end{abstract}

%\pacs{98.70.Sa, 13.85.Tp, 98.65.Dx, 98.54.Cm}
\end{frontmatter}

%\maketitle

\section{Introduction}

The cosmic microwave background is expected to limit the travel
distance of nucleons and nuclei above 30~EeV, due to
photo-disintegration or photo-production of pions.  These interactions
impose a cut-off in the spectrum of high-energy cosmic rays known as
the GZK cutoff~\cite{gzk}, strongly limiting their visible flux.

The usual interpretation of the expected reduction of the flux above
100 EeV lies in the sudden reduction of the visible Universe from a
few Gpc below 10 EeV to less than 20 Mpc above a few 100 EeV. This
simple view may however be strongly modified in the presence of
extra-galactic magnetic fields of the order of 100~nG. Such strong
fields are compatible with current observations and upper limits from
Faraday rotation measurements which indicate field strengths at the
$\mu$G level within the central Mpc region of galaxy
clusters~\cite{bo_review,ryu,blasi,kronberg}.  Relatively strong
magnetic fields may exist in intergalactic space with coherence
lengths of the order of a Mpc.

In realistic models of the Universe, however, such strong fields
cannot fill the entire intergalactic volume, and should rather be
concentrated along the high density sheets where matter also
concentrates, as suggested by large-scale structure formation models
\cite{sigl}.  The transport of UHECRs in this kind of environments, and
their resulting spectrum observed on Earth, depend on the particular
geometry of the fields, combined with that of the sources, and on the
intensity of the magnetic fields.  This has been investigated by a
number of authors using various assumptions for the distance of one or
a few local sources, and working out UHECR propagation in a model of
the local supercluster \cite{WW,G33,G34,G36,BO,med-tan}.  It has been 
found that under some circumstances, the UHECR spectrum could show 
an attenuated GZK suppression.

In this paper, we focus on the effect of strong magnetic fields around
our Galaxy, assuming that the local group lies within a supercluster
with higher than average magnetic field.  In order to obtain a clear
physical understanding of this effect alone, we use a spherically
symmetric configuration and a homogeneous field in the Universe.  This
is clearly not representative of the observed walls-and-voids
structure of our Universe on all scales, but it allows us to emphasize
a basic mechanism using a complete Monte-Carlo propagation model, and
disentangling the configuration specific effects from the influence of
strong fields.

As a matter of fact, in magnetic fields of a 100~nG, a charged
particle of 10~EeV has a Larmor radius of 100~kpc and propagates
diffusively while at higher energies (above 100~EeV, say) the
trajectories are essentially ballistic.  Therefore the clear picture
of the GZK cutoff imposed by particle dynamics gets blurred by another
effect taking place at the same energy: the transition between
ballistic and diffusive transport regimes.

Another point to note is that in such a field and for attenuation
times of the order of a few Gyr (attenuation of pre-GZK protons by
pair production), the radius of the diffusion sphere of a continuous
source is only 10~Mpc.  Therefore, far away sources ($D \geq 100$~Mpc,
say), would not be visible at any energy.

Given the above ingredients, nearby sources ($10~\mathrm{Mpc} \leq D
\leq 100$~Mpc) appear to be responsible for the observed UHECR spectrum
at \textit{all} energies.  These sources would be sufficiently far
away so that we lie outside their low-energy (1--10~EeV) diffusion
spheres, therefore reducing the corresponding flux, but close enough
so that their high-energy component does not get significantly
attenuated by energy losses.  This can reduce the usually inferred
flux difference between energies below and above the GZK energy.  In
addition, on a 10~Mpc scale the Universe cannot be taken as isotropic
and uniform, which indicates that the local distribution of matter
will play a dominant role in the actually observed energy spectra.

In the following we explore this scenario.  The next section describes
our magnetic field model and our particle propagation Monte-Carlo.
Section 3 discusses the magnetic horizon for charged particles and
section 4 presents our results.

\section{A simplistic magnetized Universe}
\subsection{Field modelling}
We simulated a magnetized Universe following the method
of~\cite{O49,O14}.  Our turbulent field has a zero mean value and
fluctuates following a Gaussian distribution.  The spectrum of
fluctuations is a power law \[ <\delta B(\vec{k})^2> = S_{0}
k^{-2-\beta} \]
The normalization $S_{0}$ is obtained through the
Wiener-Kintchine theorem by imposing the average fluctuation strength
$<\delta B^2>$: \[ <\delta B^2> = \int{S_0k^{-2-\beta}4\pi
k^2\mathrm{d}k} \]
leading to \[ S_0 = \frac{\beta-1}{k_m^{1-\beta}-k_M^{1-\beta}} <\delta
B^2> \]
The maximum ($k_M$) and minimum ($k_m$) modes are set respectively
by the size of the simulation box $L_c=$1 Mpc
(related to the coherence length of our field model)
and the step size of the grid $l=$15 kpc.  We used a
Kolmogorov fluctuation spectrum corresponding to $\beta=5/3$.  In
Fourier space, the field verifies $\vec{B_k}.\vec{k}=0$ to satisfy
div$\vec{B}=0$ and we pick up a random phase in the transverse plane
to complete the field description.  In real space the field is
computed on a 15~kpc grid over a volume of 1 Mpc$^3$.  This volume is
then translated periodically to fill our modelled Universe.  The value of
the field at any point in space is obtained through a linear
interpolation of the 8 closest vertices of the grid.

\subsection{Particle trajectories and sources spectra}

For each Monte-Carlo simulation, we compute a field configuration according 
to the above method and we follow a number of protons as long as their
energy is above 10~EeV or their propagation time is below 3~Gyr.  At
each step along their trajectory, we solve the equation of motion in
the local field and compute the energy loss due to pion
photo-production, simulating each interaction. The losses due to pair
production are treated continuously.  Neutrons, if produced, are also
followed until they transform again into a proton via the
photo-production of pions or until they decay.

To compute the spectrum of a source at a given distance $R$ from the
observer, we generate a set of protons at the origin and record the
times (and the corresponding energies) of each crossing of a sphere of
radius $R$, centered on the origin.  By construction, our spectrum does
not depend on the crossing position on the sphere, therefore we record
them all.  Moreover the effective detection probability being very
small (the surface of a detector is negligible compared to the surface
of the sphere), we compute the spectrum at various radii following the
same set of particles.

\begin{figure} [!t]
%\vspace*{-1cm}
\centering\includegraphics[bbllx=30,bblly=80,bburx=535,bbury=310,clip=,width=11cm,height=7cm]{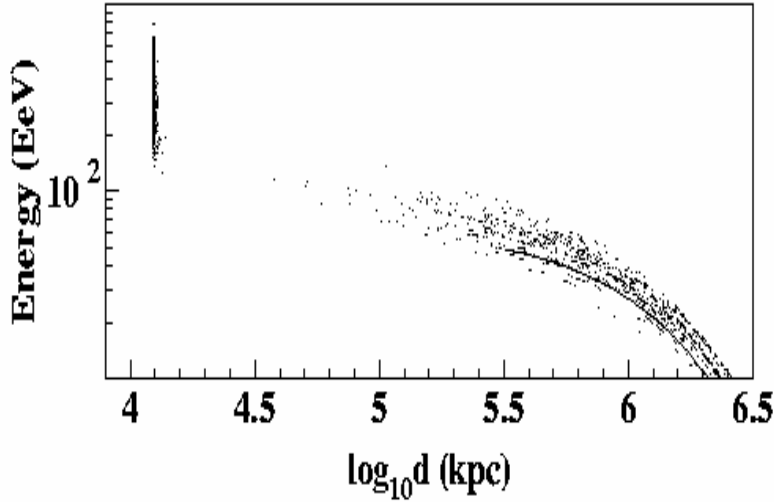}
%\centering\epsfig{file=trans_100ng_800.eps,width=8.5cm,height=7.5cm}
%\vspace*{-1cm}
\caption{Particle energies as a function of the distance of travel in
a 100~nG field for an observer lying on a sphere 12~Mpc away from the
particle source.  The initial energy is 800 EeV. On the left, an
accumulation of points represents the particles that did not interact
and follow a ballistic trajectory; on the right, the solid line
represents the expected contribution of the continuous pair-production
energy losses.}
\label{trans}
\end{figure}

Fig.~\ref{trans} shows the distribution of particle energies as a
function of trajectory length (equivalent to time) for an observer
located on a sphere 12~Mpc away from the source, and for an initial
energy of 800~EeV.

For radii comparable to or smaller than $L_c$ the orientation of the
field on the surface of the sphere may not average properly and may
lead to strong distortion in the time distribution of the particle
crossings (a particle trapped along a field line parallel to the
sphere surface may cross it many times over a small distance).  To
avoid this effect we regenerate a field configuration every 100
generated protons.

From the recorded information we are able to compute the probability
$\mathcal{F}(t-t_0,E;R,E_0)\ud E\ud t$ for a particle produced at time
$t_0$ with energy $E_0$ to be detected at a given distance $R$ at time
$t$ with an energy $E$.  From these probability tables, and assuming
an isotropic distribution of sources $\rho(R) = \rho_{0}$, the
spectrum observed today (time t) from sources located between
$R_{min}$ and $R_{max}$ is given by:
\begin{eqnarray}\label{eqn:flux}
\frac{\ud N}{\ud E}(t) = &\rho_{0}\int_{R_{min}}^{R_{max}} \ud V
\int_0^t\ud t_0\int_{E_0^{min}}^{E_0^{max}}\ud E_0
f(E_0) \frac{1}{4\pi R^2} \mathcal{F}(t-t_0,E;R,E_0)\,,
\end{eqnarray}
where $f(E_0)$ is the source injection spectrum and where we have
neglected all cosmological effects.

\section{Magnetic horizon}
As we mentioned in the introduction, for field strengths of the order
of 100~nG with Mpc coherence length, particle trajectories below a few
tens of EeV are well described by the diffusion equation. Several
diffusive regime may be distinguished within a given field
configuration, depending on the particle energy. For the pure random
field we are considering, three regimes have been identified and can
be parameterized with the following diffusion coefficients (in
Mpc$^2$/Myr)~\cite{O14}:

\begin{eqnarray}\label{eqn:DE} D(E) & \simeq & 2\,10^{-2}
\Big(\frac{E}{B}\frac{\mu\textrm{G}}{10^{20}\textrm{eV}}\Big)^{7/3}
\Big(\frac{L}{\textrm{Mpc}}\Big)^{-4/3}
(E > E^*)\nonumber\\
& \simeq & 3\,10^{-2}
\Big(\frac{E}{B}\frac{\mu\textrm{G}}{10^{20}\textrm{eV}}\Big)
\,\,\,(0.1 E^* < E < E^*)\\
& \simeq & 4\,10^{-3}
\Big(\frac{E}{B}\frac{\mu\textrm{G}}{10^{20}\textrm{eV}}\Big)^{1/3}
\Big(\frac{L}{\textrm{Mpc}}\Big)^{-2/3}
(E < 0.1 E^*)\nonumber
\end{eqnarray}

\begin{figure}[!t]
%\hspace*{-0.3cm}
\centering\epsfig{file=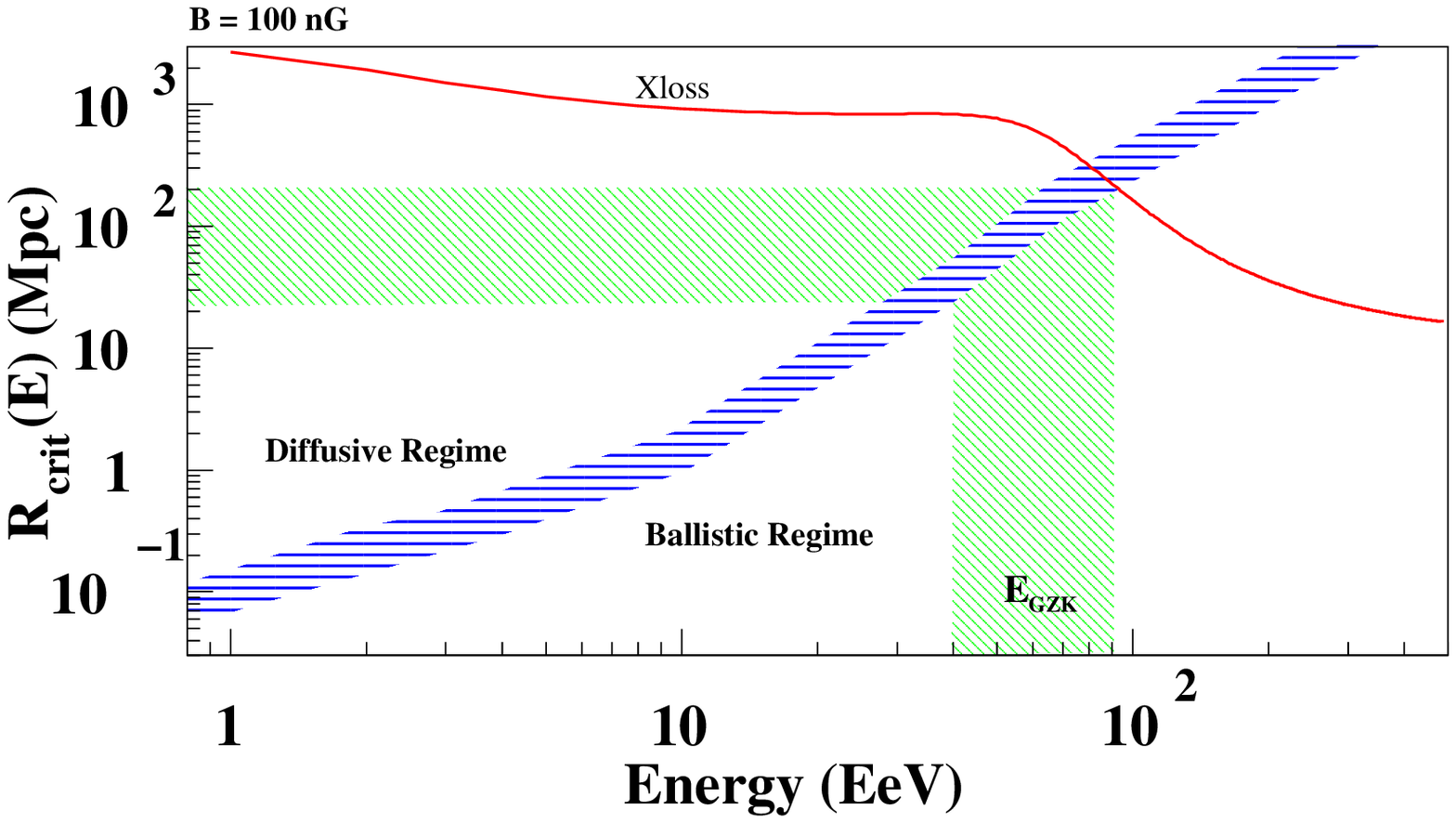,width=11cm,height=7cm}
\vspace*{-0.3cm}
\caption{Regions in the plane $(E-R)$ where
ballistic or diffusive regime describes particle trajectories for a
100 nG field. The transition occurs around a
few tens of Mpc, i.e. inside the GZK sphere at GZK energies.}
\label{fig:trans_ball_diff}
\end{figure}

The transition energy $E^*$ can be evaluated from the Larmor radius and
the field coherence length $r_L(E^*) = L_c/2\pi$, giving: \[ E^* \simeq 
1.45\times10^{20} \Big(\frac{B}{\mu\textrm{G}}\Big)
\Big(\frac{L}{\textrm{Mpc}}\Big) \textrm{eV}
\]
Of course, the transition actually takes place over a range of 
energies around $E^*$.

Below the GZK cutoff energy, where losses do not play a crucial role,
the Green function for the diffusion is given by: \[ n(E,\vec{r},t) =
\frac{N_0}{(8\pi D(E)t)^{3/2}} \exp{\Big(-\frac{r^2}{4D(E)t}\Big)}
\]

From this solution, one can estimate the radius of the diffusion
sphere as a function of time and energy: \[ R(E,t) \sim \sqrt{4D(E)t}
\]

For a given source distance $R$, the energy scale separating the
ballistic and diffusive regimes can be estimated by comparing the
corresponding propagation times: $R^{2}/4D(E_{\mathrm{c}})$ and $R/c$.
Fig.~\ref{fig:trans_ball_diff} shows this energy as a function of
source distance for a random field of 100~nG. It is remarkable that
for sources within the GZK sphere (10--100~Mpc), the transition occurs
around the GZK energy (i.e. a few tens of EeV).

The existence of a diffusion sphere whose radius grows much slower
than $ct$ gives rise to a \emph{magnetic horizon} which limits the
distance up to which a given particle can escape from its source
before losing most of its energy.  This horizon is given by:
\begin{equation}\label{eqn:horizon}
  R_H = \sqrt{4D(E)T_{loss}(E)}
\end{equation}
where $T_{loss}(E)=E/\frac{\ud E}{\ud t}$ is the energy loss time.

\begin{figure}[!t]
\centering\epsfig{file=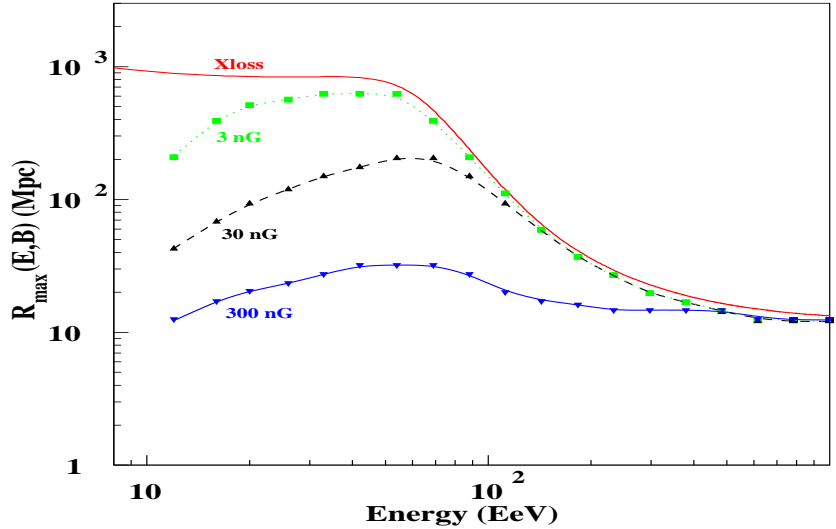,width=11cm,height=7cm}
\caption{Magnetic horizon as a function of energy for various field
strengths.  The horizon is defined as the radius of the sphere
enclosing 68\% of the particles trajectories.  The maximum propagation
time is fixed by the energy loss time (top solid line). One can see that
for a 300 nG random field, the horizon is fairly constant around 20 Mpc 
at all energies.}
\label{fig:horizon} \end{figure}

Using the Monte Carlo described in the previous section, we computed
this distance as the maximum radius reached by 68\% of the particles
over a propagation time $T_{loss}(E)$. This horizon is shown on
Fig.~\ref{fig:horizon}. At lower energy (below 10 EeV), it is given by
Eqn.~\ref{eqn:horizon} and the parameterization of the diffusion
coefficients given in Eqn.~\ref{eqn:DE}. Different values of turbulent 
magnetic fields have been used. For relatively strong fields (300 nG), 
the horizon is the same before and after the GZK energy ; implying that
the sudden reduction of the visible Universe at the origin of the 
expected suppression of the flux above 100 EeV does \emph{not} occur 
in that case.

\section{Results}
All the following results have been obtained for continuous sources
with power-law injection spectra of index 2.3 and maximum energy of
1000~EeV. This slope is typical of particle acceleration at
relativistic shockwave~\cite{ell-dou,ach-gal,lem-pel}.

\begin{figure}[!t]
%\centering\includegraphics[bbllx=30,bblly=80,bburx=535,bbury=540,clip=,width=12cm,height=7cm]{300ng_all-sources2.eps}
\centering\epsfig{file=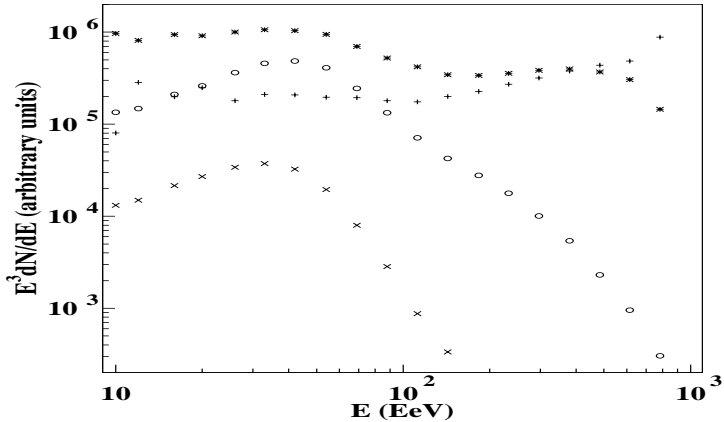,width=11cm,height=7cm}
\caption{Comparison of spectra obtained for sources located at 1 ($+$),
10 ($\star $), 50 (o) and 100 ($\times $) Mpc in a turbulent field of 300nG. The
solid angle effect has been taken away, so that the normalization of
the four spectra would be identical in the absence of magnetic field and
energy losses.  As expected, the GZK cutoff is visible for far away
sources, but the latter have a negligible contribution to the total flux at
all energies.}
\label{fig:4sources}
\end{figure}

We calculated the observed spectra for various source positions (45 values
distributed uniformly in log scale from 1~Mpc to 1~Gpc) and turbulent field
values ranging from 3 to 300 nG.  Fig.~\ref{fig:4sources} shows a comparison of
these spectra for sources at 1, 10, 50 and 100 Mpc obtained with a 100 nG
random field.  The solid angle effect has been taken away so that the
normalization and spectra at all distances would be identical in the absence of
magnetic field and energy losses.  As expected, the spectrum of a very close
source (1 Mpc) shows a softening (index changing form $-2.3$ to about $-3.0$)
at low energies due to the accumulation of low energy particles, while at high
energy the original spectrum is restored.  For such a nearby source there is of
course no GZK cutoff, but this configuration would lead to an anisotropy
in the sky incompatible with the already available data.  At intermediate 
distance (10 Mpc) the flux is only slightly reduced at high energy and the 
contribution (corrected by E$^3$) at 10 EeV and 1000 EeV are roughly the 
same, leading to a nearly flat spectrum in the 10--1000~EeV range.

Further away (above 100 Mpc) the flux is strongly attenuated at high energy
because of the GZK cutoff but also at low energy because of the diffusive
regime.  One should also note that increasing the distance between the source
and the observer leads to a stronger GZK cutoff because the mean travel length
increases faster than the source distance. Moreover, for greater source
distances, the diffusive regime extends to higher energy, as shown by
Fig.~\ref{fig:trans_ball_diff}.

The main conclusion is therefore that far away sources have a negligible
contribution to the total flux at \emph{all} energies (and not only above GZK
energy).  In other words, the presence of random magnetic fields leads to a non
trivial configuration of sources contributing to the spectrum, even in the case
of a uniform distribution.

\begin{figure}[!t]
%\centering\epsfig{file=figure/figure/bal_300ng_10mpc.eps,width=9cm}
\centering\epsfig{file=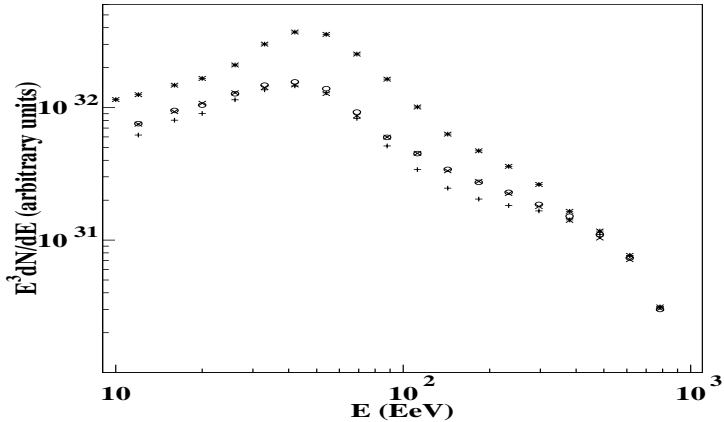,width=11cm,height=7cm}
\caption{Spectrum for a uniform distribution of sources between 10 and 1000
Mpc without random magnetic field ($\star$) and in the presence of a
300 nG field ($+$). In the later case the low energy part of the
spectrum is reduced leading to a flattening of the GZK cutoff. Also shown are 
the spectra ($o$ and $\times $) corresponding to non uniform fields, with
alternated voids and sheets regions (see text).}
\label{fig:spectrum}
\end{figure}

In Fig.~\ref{fig:spectrum} we compare the observed flux for a uniform
distribution of sources between 10 and 1000 Mpc with and without a 300~nG
random magnetic field.  In the later case the reduction of the low energy flux
is clearly visible as the observer lies mostly outside the diffusion spheres.
The exact form of the spectrum strongly depends on the source configuration on
a scale (a few tens of Mpc) where the Universe cannot be taken as uniform.
Therefore the local configuration of sources may lead to a spectrum that lies
anywhere between the shapes of those obtained from individual sources at 10 or
50 Mpc as shown on Fig.~\ref{fig:4sources}.

This comparison also shows that for a field of 300 nG, propagation
becomes purely ballistic only above 500 EeV and the transition between
the diffusive and ballistic regimes spreads over half a decade above
100 EeV. This transition implies a flattening of the spectrum at
energies higher than $E_{GZK}$.  The position of this flattening
strongly depends on the position of the most contributing sources, and
on the value of random fields.  Increasing the value of the random
fields ($\geq 3 \mu$G) leads to particle diffusion on small scales (a
few tens of Mpc) even at energies higher than $E_{GZK}$, and therefore
moves the transition to the ballistic regime to energies so high ($>
10^{21}$ eV) that this transition becomes invisible.  In such a
situation a strong GZK cutoff is restored as the signature of the much
shorter energy loss time of post GZK particles.

\section{Conclusion}

We have modelled a fully magnetized Universe to study the effect of
random magnetic fields on the spectra of UHECR sources, in a
spherically symmetric Universe, in order to get rid of any additional
effect related to the geometry of magnetic fields and source
distributions.  We showed that for magnetic fields in the 100~nG
range, sources more distant than about 100 Mpc do not contribute
to the observed fluxes at \emph{all} energies.  In such a model, the
argument of the GZK cutoff in its original form in terms of a
sudden reduction of the horizon is modified.

We are aware that a fully magnetized Universe is not a realistic model
of our environment.  Voids are known to exist, where random fields
would rather be around a few~nG, or more less. In such regions, UHECRs propagate
ballistically even at low energies where GZK losses are unimportant,
and therefore the particular effect which we discussed above does not
appear.  However, our main argument that \emph{"far away sources do
not contribute to the visible spectrum of UHECRs at all energies"}
would be maintained if all sources are surrounded by a random field of
order 100~nG over distances of several Mpc, as expected for sources
that lie within galaxies or clusters.

By using a locally homogeneous Universe, we implicitely restricted
ourselves to one sheet of magnetic fields and neglected the
contribution of other high-density sheets that may lie in our
neighbourhood.  While this is a direct consequence of our choice to
isolate the `magnetic cutoff' effect from any other effect influencing
the shape of the spectrum, it may also be qualitatively justified by
the fact that the highest energy particles would not be able to travel
the distances between two neighbouring sheets without losing some
energy, and the lowest energy particles would be largely reflected as
they approach the strongly enhanced magnetic fields in our own sheet,
resulting in a reduced flux contribution.  Such an (imperfect)
confinement of UHECRs between two magnetic walls contributes to
isolate regions of high magnetic fields from one another.

To illustrate the effect of non uniform fields, with alternating voids
and sheets, we also investigated models in which the fields are
concentrated in gaussian concentric shells of width $\sigma =
5\,\mathrm{Mpc}/\sqrt{2}$ (as in the description of planar magnetic
sheets in Ref.~\cite{G33}) or $\sigma = 2\,\mathrm{Mpc}/\sqrt{2}$,
and radii increasing by steps of 30~Mpc, respectively 15~Mpc.  As
shown in Figure~\ref{fig:spectrum}, the reduction of the low energy
part of the spectrum (between $10^{19}$ and $10^{20}$~eV) is still
visible in this field configuration, and the general effect of
reducing the GZK suppression remains.  More realistic field
configurations thus do not seem to affect the general arguments
emphasized here.

We did not model any particular configuration of sources to study the
anisotropy effect, but given the slowness of the transition between
the two propagation regimes and the fact that the transport becomes
only purely ballistic above 500~EeV, we expect that anisotropies would
be visible only above 100~EeV where the data is currently too scarce
for one to make any definitive statement.  It has been shown
recently that even with more localized
magnetic fields the expected anisotropy from local sources is not
ruled out by the current data~\cite{sigl}.  One should also note that a few
sources localized in the magnetic sheet around the local supercluster
would suppress the bump before $E_{GZK}$, visible on
Fig.~\ref{fig:spectrum}, because low-energy particles leak through the
walls~\cite{G33}.

% Finally we want to stress that this work was done to emphasize a
% mechanism and is therefore based on a simplified model of a magnetized
% Universe.  While doing this study we went in detail through the
% numerous literature on the subject~\cite{G33,G34,G36,G37,BO} (and
% references therein).  We realize that some of our arguments and
% conclusions are spread in these papers, however not brought together
% nor summarized in a simple model as we did here.

\section*{Acknowledgments}
We would like to thank Ricardo Perez for invaluable discussions.

\end{document}